**AI-powered virtual eye: perspective, challenges and opportunities**

Yue Wu[1#], Yibo Guo[2#], Yulong Yan[3], Jiancheng Yang[4], Xin Zhou[5,6,7], Ching-Yu Cheng[8,9], Danli Shi[1,10,11*], Mingguang He[1,10,11*]

**Affiliation**

1. School of Optometry, The Hong Kong Polytechnic University, Hong Kong SAR, China

2. Key Laboratory of Carcinogenesis and Cancer Invasion, Liver Cancer Institute, Zhongshan Hospital, Fudan University, Shanghai, China

3. School of Medicine, Shanghai Jiao Tong University, Shanghai, China

4. Swiss Federal Institute of Technology Lausanne (EPFL), Lausanne, Switzerland

5. Intelligent Medicine Institute, Fudan Microbiome Center, Fudan University Shanghai Medical College, Fudan University, Shanghai, China.

6. Department of Genetics, Stanford University School of Medicine, Stanford, CA, USA

7. Collaboratory on Longitudinal Deep Omics, The Hong Kong Polytechnic University, Hong Kong SAR, China

8. Department of Ophthalmology, Yong Loo Lin School of Medicine, National University of Singapore, Singapore, Singapore

9. Singapore Eye Research Institute, Singapore National Eye Centre, Singapore, Singapore

10. Research Centre for SHARP Vision (RCSV), The Hong Kong Polytechnic University, Hong Kong SAR, China

11. Centre for Eye and Vision Research (CEVR), 17W Hong Kong Science Park, Hong Kong SAR, China

#Contributed equally

**Correspondence**

Prof. Mingguang He, MD, PhD., Chair Professor of Experimental Ophthalmology, School of Optometry, The Hong Kong Polytechnic University, Hong Kong, China. Email: mingguang.he@polyu.edu.hk

Dr. Danli Shi, MD, PhD. The Hong Kong Polytechnic University, Kowloon, Hong Kong SAR, China. Email: danli.shi@polyu.edu.hk.

**Abstract**

We envision the "virtual eye" as a next-generation, AI-powered platform that uses interconnected foundation models to simulate the eye's intricate structure and biological function across all scales. Advances in AI, imaging, and multi-omics provide a fertile ground for constructing a universal, high-fidelity digital replica of the human eye. This perspective traces the evolution from early mechanistic and rule-based models to contemporary AI-driven approaches, integrating in a unified model with multimodal, multiscale, dynamic predictive capabilities and embedded feedback mechanisms. We propose a development roadmap emphasizing the roles of large-scale multimodal datasets, generative AI, foundation models, agent-based architectures, and interactive interfaces. Despite challenges in interpretability, ethics, data processing and evaluation, the virtual eye holds the potential to revolutionize personalized ophthalmic care and accelerate research into ocular health and disease.

# 1. Introduction

A computational eye model aims to simulate, generate, predict, and analyze the structural and functional states of the eye. Owing to its rich imaging landscape and well-characterized anatomy, the eye serves as an ideal organ for virtual reconstruction. Traditional approaches to modeling ocular processes have relied on mechanistic, rule-based frameworks driven by mathematical formulations and biological priors, enabling simulations of biomechanical dynamics, optical imaging, and pharmacokinetics[1-5]. While instrumental in advancing our understanding of disease mechanisms, such models are typically limited in scope, often tailored to narrow questions and constrained in their ability to integrate multi-source data or simulate dynamic, cross-scale behaviors.

Recent breakthroughs in artificial intelligence (AI) have opened new horizons for developing a next-generation virtual eye. Inspired by pioneering efforts in the creation of virtual cells and hearts[6-8], the AI-powered virtual eye is conceptualized as a platform of interconnected foundation models that capture biological dynamics across multiple levels of abstraction. In contrast to earlier models focused on single-task applications, the virtual eye aspires to be a comprehensive and holistic system, capable of "seeing, interpreting, and predicting" with consistency across diverse clinical contexts. By combining mechanistic insights with data-driven intelligence, such a platform could bridge the gap between theory and empirical data, thereby supporting precision diagnostics, treatment planning, and personalized medicine in ophthalmology.

This review presents a forward-looking perspective on the development of the AI-powered virtual eye. We begin with a historical overview of eye modeling, then introduce a roadmap for the virtual eye's construction, including key enablers, technical challenges, and prospective applications. Our aim is to provide a comprehensive synthesis of the current landscape while highlighting the transformative potential of this technology in biomedical research and clinical care.

# 2. Conceptual evolution of the eye model

The pursuit of a "virtual eye" began with early computational models that used mathematics, physics, statistics, and computer science to simulate ocular systems. These models incorporated interdependent variables to enable analysis of how perturbations affect ocular function and system performance[9]. The development of the virtual eye model progresses as its functionality and complexity increase. Below, we outline three critical stages that have collectively shaped the conceptual architecture of the virtual eye:

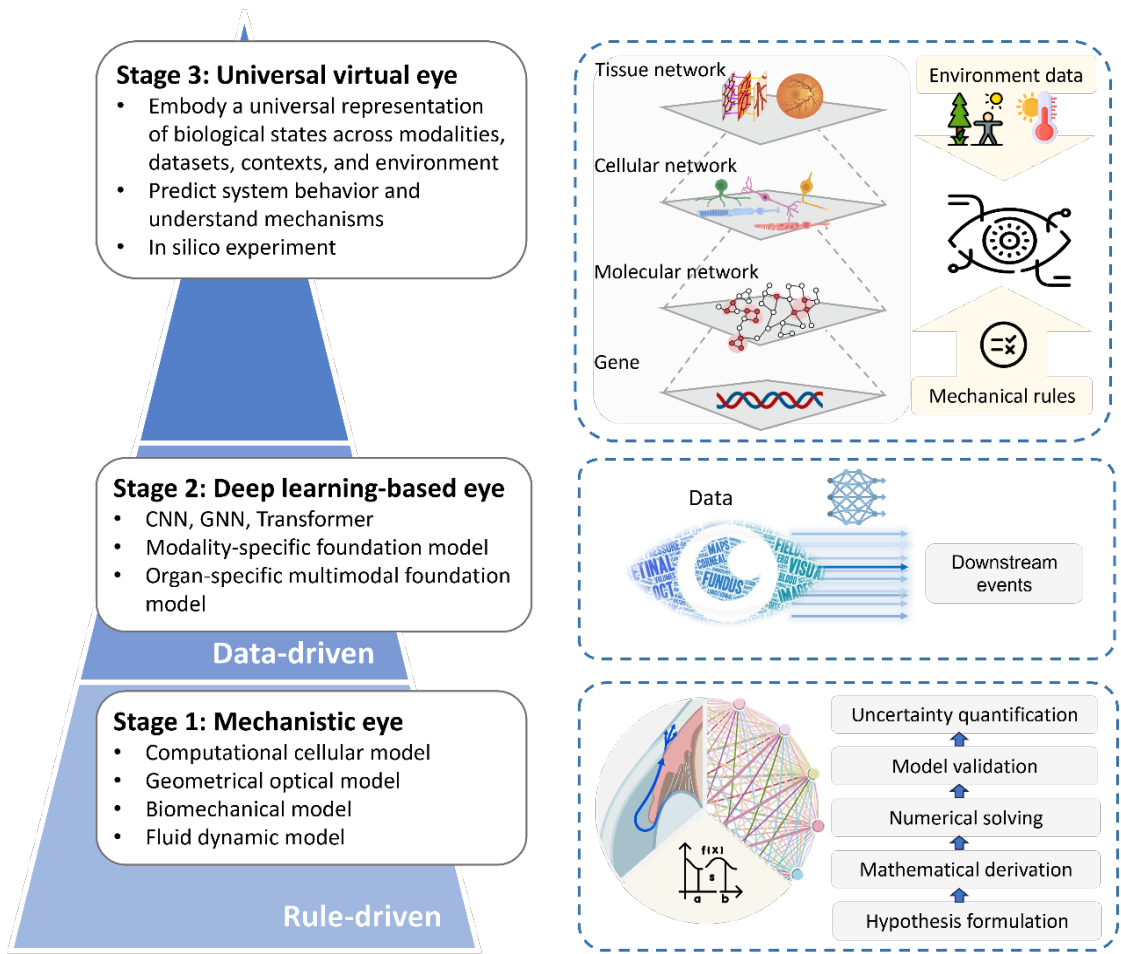

**Figure 1. Evolutions of the virtual eye.**

## 2.1 Stage 1: Mechanistic eye model

Early computational models were mechanistic in nature, grounded in established anatomical and physiological knowledge. In these models, each ocular component, cornea, lens, retina, and others, was represented mathematically to simulate behavior under defined conditions. As summarized in **Table 1**, such models span multiple biological levels. At the molecular scale, they simulate key processes such as protein-protein interactions, enzymatic reactions, signaling pathways, and ion dynamics. At the organ level, geometrical optical models ranging from Gullstrand simplified eye model[10] to more advanced simulations using platforms like Zemax[11], have enabled the predictions of retinal image size, refractive errors, and the effects of optical interventions such as contact lenses[12-14], intraocular lenses[15,16], and refractive surgeries[17]. Biomechanical models have further supported investigations of stress-strain responses and tissue deformation under internal and external forces, in processes such as corneal expansion after refractive surgery[18], myopic scleral remodeling[19], and optic nerve damage in glaucoma[20,21]. Fluid dynamics analyses have also been employed to study aqueous humor circulation[22,23], retinal hemodynamics[24], tear film dynamics[25], and drug distribution kinetics[26]. Importantly, patient-specific structural data can support both fluid dynamics and biomechanics analysis. For example, recent work has used three-dimensional retinal vasculature reconstructions from OCT angiography not only to simulate structural-based dynamics but also to perform fluid-structure interaction simulations, comparing the induced tissue stresses in diabetic and healthy conditions [27].

Models in this stage typically follow a bottom-up approach, beginning with a specific biological question, incorporating simplified assumptions, and using mathematical solutions validated against empirical data. While they offer valuable causal insights, they are often constrained by reliance on population-averaged parameters and limited generalizability beyond predefined physiological conditions.

**2.2 Stage 2: Deep-learning-based eye model**
AI-based eye models mark a paradigm shift from rule-based systems to data-driven frameworks. Unlike mechanistic models, AI models do not rely on explicitly defined physical equations. Instead, they use machine learning, particularly deep learning, to uncover latent relationships between complex input data and clinical outcomes.

These models are often designed to reduce human intervention and are particularly useful in scenarios where biological mechanisms are unclear or incompletely characterized. The structure of these models is influenced by the nature of the input data and the clinical task at hand (see **Table 1**). Many current models project multimodal inputs into shared latent spaces, enabling the learning of cross-modal correlations and supporting predictions of downstream effects from changes in input variables.

A milestone was the emergence of foundation models like RETFound, which was pre-trained on millions of fundus images and can be fine-tuned for diagnostic tasks in a data-efficient manner[28]. By providing a generalizable backbone rather than a narrow single-purpose network, Foundation models achieved high accuracy in disease detection with minimal retraining. Multimodal foundation models like EyeFound, VisionFM and EyeCLIP, further expanded the ophthalmic modalities to learn a unified image representation, representing an early form of an organ-specific foundation model[29-31].Compared to mechanistic models, multimodal foundation models offer superior scalability and are capable of handling large, heterogeneous datasets. However, they often sacrifice interpretability, lack explicit causal modeling, and remain vulnerable to distributional shifts across datasets. Furthermore, while they excel in specific visual tasks, current AI models remain task-specific and have yet to achieve seamless integration across cellular to organ-level functions.

**2.3 Stage 3: Towards a Universal Virtual eye**

The next generation of eye models aspires to synthesize the strengths of both mechanistic and deep learning approaches to create a universal virtual eye. Drawing inspiration from Bunne et al.'s AI virtual cell framework[6], this stage envisions a general-purpose representation of the human eye that integrates physiological knowledge with data-driven learning across scales, modalities, and contexts.

Here we envision the universal virtual eye as a comprehensive AI framework composed of interconnected foundation models capable of representing biological structure and function with high fidelity. The universal virtual eye should exhibit the following key characteristics: (1) multi-modal modeling capability; (2) multi-scale integration; (3) representation of diverse and dynamic process; and (4) incorporation of complex feedback loops **(Table 2, Figure 2)**

**2.3.1 Multi-modal modeling capability**: The virtual eye will leverage diverse data sources to construct a holistic representation of ocular systems. . These include: 1) structural and functional imaging (e.g., optical coherence tomography (OCT), color fundus photography (CFP), fluorescein fundus angiography(FFA), electroretinogram (ERG), visual field(VF)), 2) molecular profiles (e.g., genomics, proteomics, metabolomics), 3) clinical data (e.g., electronic health records, longitudinal phenotypic data, comorbidities, and surgical history), and 4) environmental data (e.g. light exposure, ambient temperature, and other contextual factors).

**2.3.2 Multi-Scale Integration:** The model will integrate biological processes across spatial and temporal scales from nanoscale molecules to the macrostructure of the eye. At the molecular level, it will simulate gene regulatory networks and predict how genetic variants influence protein function. At the cellular level, it will capture signaling and metabolic dynamics, linking them to higher-order tissue behaviors. Ultimately, the system will bridge microscopic events with macroscopic clinical outcomes.

**2.3.3 Representation of diverse and dynamic process:** The virtual eye will not only replicate known biological behaviors and predict responses to novel interventions, but also capture temporal dynamics. This enables time-resolved simulations of critical phenomena such as development, disease progression, homeostasis, repair, and aging, making it possible to forecast future states and identify intervention points.

**2.3.4 Complex feedback loops:** The model will include both internal feedback mechanisms (e.g., gene-protein regulation, neurovascular coupling) and external feedback loops involving real-world data. In digital twin applications, the virtual eye can be continuously updated using patient-specific data, allowing for dynamic recalibration and adaptive learning. This dual feedback system ensures both biological coherence and responsiveness to real-world inputs.

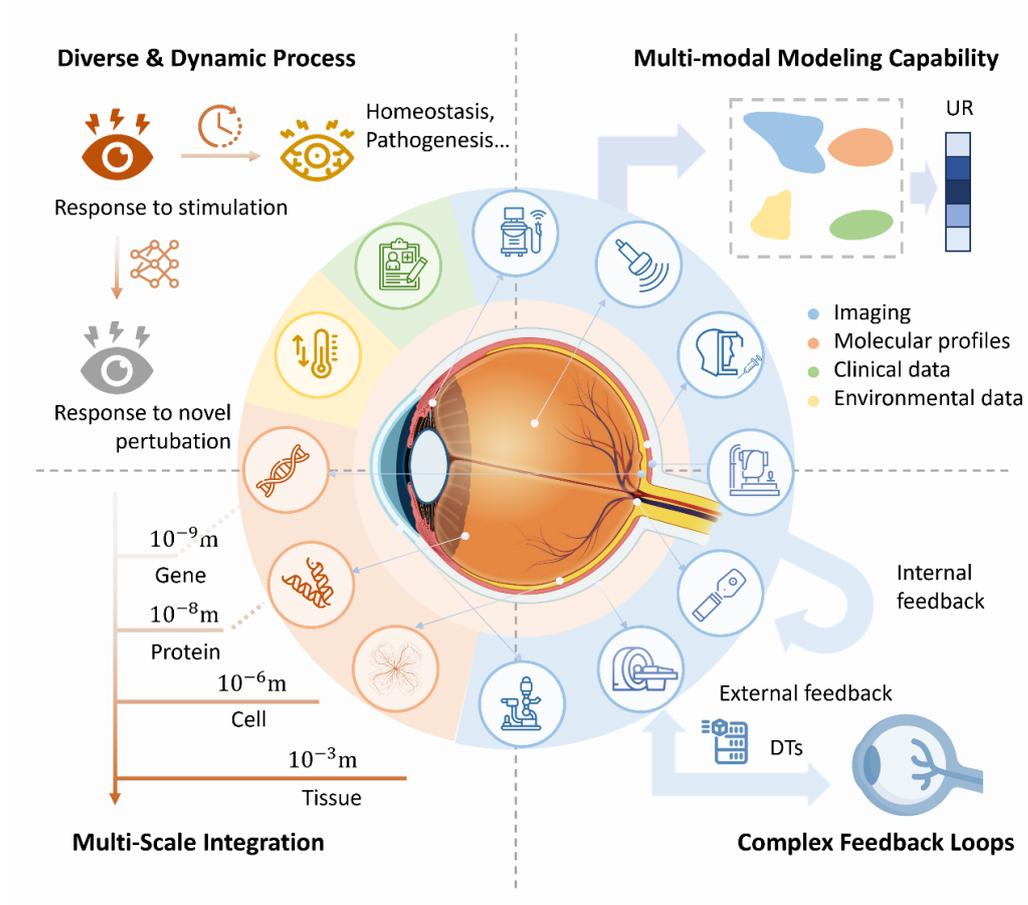

**Figure 2. Hallmarks of the universal Virtual eye**

**3. Roadmap for virtual eye with AI: data, modeling, and interaction**
Building the virtual eye with AI is an interdisciplinary system engineering challenge. To properly model such complex behaviors, many approaches should be explored and their merits carefully judged. Here, to better articulate the technical details of its construction, we describe these sections focusing on data acquisition, processing and interaction with human and environment.

**3.1 Data**

**3.1.1 Multimodal and large-scale dataset**
At the heart of the virtual eye is the integration of heterogeneous data into a unified, spatiotemporally aligned reference framework. Recent advances in imaging and sensing technologies significantly improve the resolution, scale, and depth of ocular data. Modalities such as ultra-widefield 3D OCT, adaptive optics scanning laser ophthalmoscopy (AO-SLO), and optical coherence elastography (OCE) expand our ability to characterize ocular structures and biomechanical properties with unprecedented detail[32,33]. Single-cell genomics reveals cellular heterogeneity[34], while spatial multi-omics maps molecular signals to 3D tissue structures[35], offering an intricate view of ocular microenvironments. Concurrently, continuous environmental and physiological data streams from electronic health records, smart contact lenses[36,37] and wearable devices[38,39], transforming a static dataset into a dynamic, ever-evolving record. This ongoing flow of personalized information enables virtual eye to track and adapt to individual disease trajectories in real time, improving predictive modeling and early intervention.

**3.1.2 Generative AI for Synthetic Data**
To address real-world data scarcity and enhance training diversity, generative AI techniques such as variational autoencoders (VAEs), GANs, diffusion models, and autoregressive models are being applied to synthetic data generation across multiple tasks. These include image-to-image/video, text-to-image/video, and 3D structure generation. For instance, CFP can be transformed into FFA or indocyanine green angiography (ICGA) images, reducing the need for invasive diagnostics[40,41]. Diffusion-basedand SORA-like models generate 2D ophthalmic images and videos for education and diagnosis[31,42], whereas systems like ChromoGen[43], AlphaFold[44,45], Rosetta Fold[46], and Fundus2Globe[47] demonstrate the feasibility of reconstructing 3D structures (chromatin, protein, and eye shape) from amino acid sequences, DNA sequences, and planar imaging data. Incorporating reinforcement learning, synthetic-real data comparisons can drive a feedback loop for continual refinement, transforming imaging from a passive observation tool into an active, generative component of virtual eye development[48].

**3.2 Modeling: architecture and downstream tasks**
**3.2.1 General foundation model vs AI agent**
Achieving unified multi-modal, multi-task modeling remains an open challenge, primarily due to substantial heterogeneity across data modalities, uneven data scales, and complex cross-scale requirements. Although some recent efforts employ self-supervised learning to align unlabeled multimodal features, for example, CFP phenotypes and genetic feature[49]; they have limited capacity for cross-scale prediction and generative tasks. The emergence of foundation models offers a promising path forward: large-scale cross-modality pretraining and contrastive learning can produce a shared representation[50], and it has the ability to balance extended contextual information with fine-grained sensitivity. EVO, a foundation model developed by Nguyen et al., captures the inherent multi-modality and multi-scale evolutionary features of the central dogma[51]. This unifies different data modalities (DNA, RNA, and protein) into a single codified, predictive information stream. Despite advancement in molecular, cellular[52] and tissue-level foundation model[28,30], they largely operate independently and a unified framework linking molecules, pathways, cells, and whole organ remains elusive.

A practical starting point may involve building modality-specific foundation models, then linking complementary ones through modular pipelines. In this design, each model outputs to a centralized decision module, or alternatively, ensemble and mixture-of-experts architectures may be employed to route tasks dynamically based on data characteristics [53-55].

A more ambitious vision involves creating adaptive AI agents that autonomously learn, reason, and generalize across domains[56,57]. Such agents could assimilate new data continuously, update internal representations in real-time, and adjust output based on evolving clinical knowledge and individual patient profiles.

### 3.2.2 Simulation and Prediction
Once a shared representation is established, task-specific modules can be fine-tuned for classification, segmentation, forecasting, or drug response prediction. A critical function of the virtual eye will be its ability to simulate future biological states. For example, recurrent neural networks or temporal convolutional networks could predict disease progression based on time-series imaging.

Beyond interpolation, the virtual eye should enable zero-shot inference—predicting the outcomes of untested interventions[58]. The MorphoDiff framework[59], which generates realistic images of cellular responses to chemical or genetic perturbations, exemplifies how generative AI can simulate "what-if" scenarios. By combining empirical data with prior knowledge, the Virtual Eye may one day model therapeutic responses before treatments are administered, enabling truly personalized medicine.

### 3.3 Interaction
For the Virtual Eye to have real-world impact, it must be accessible to clinicians, researchers, and patients alike. This requires the development of intuitive, interactive interfaces. A conversational layer powered by a domain-specific large language model (LLM) could serve as a natural access point, allowing users to ask questions, adjust variables, conduct interventions, or interpret outputs intuitively[60]. Systems trained with heterogenous data will also enable multimodal interactions, linking images, annotations, tabular data, and explanations in both directions. Additionally, embodied AI expands the Virtual Eye's capabilities by interfacing with robotics and diagnostic tools[61]. This could facilitate autonomous imaging, real-time monitoring, and even precision-guided therapies.

Together, these interaction modalities transform the Virtual Eye from a static model into a collaborative partner - an always-evolving, explainable, and actionable system for research, education, and clinical decision-making.

## 4. Challenges and recommendations

Although the virtual eye holds enormous potential, realizing its full utility requires addressing a range of technical, ethical, and practical challenges. Many of these issues are shared with traditional deep learning systems but become significantly more complex when scaled to a large, multimodal, and continuously evolving framework. Additionally, the integration of diverse and high-dimensional data sources into a unified model introduces new layers of complexity.

To navigate these challenges effectively, we recommend a "divide and conquer" strategy, in which modular subsystems are developed independently and later integrated into a cohesive Virtual Eye architecture. This modular approach allows for targeted innovation, manageable validation, and greater transparency in performance assessment. Below, we highlight key challenges and propose corresponding recommendations.

### 4.1 Model interpretability

A fundamental barrier to clinical translation lies in the lack of interpretability. As a predominantly black-box system, the Virtual Eye may obscure the rationale behind its predictions and decisions. While techniques such as SHAP, LIME, and Grad-CAM can help visualize feature contributions[62], these only partially demystify model behavior. Incorporating counterfactual reasoning and causal inference frameworks may yield deeper insights into the model's internal logic and improve trustworthiness.

### 4.2 Ethics

Ethical considerations are equally important. Models trained on non-representative populations risk introducing algorithmic bias, leading to disparities in care quality across demographic groups[63]. We recommend incorporating diversity-aware data curation, ongoing bias audits, and fairness metrics during model development. Additionally, given the sensitive nature of the biological and clinical data involved, robust data privacy protocols, secure federated learning frameworks, and transparent governance structures are prioritized[64].

### 4.3 Data redundancy and standardization

The Virtual Eye is built on vast quantities of multimodal data, yet using these data directly can lead to redundancy, noise, and inefficiency. To address this, a dedicated data-processing AI (DPAI) can be developed to autonomously annotate, clean, and harmonize heterogeneous datasets[65]. This system, powered by self-supervised learning and context-aware algorithms, can construct a unified, scalable data representation. This approach could pave the way for a common computational language that more effectively links fragmented data.

**4.4 Evaluation frameworks**

Traditional benchmarking approaches are inadequate for a system as complex as the virtual eye.[66] A more sophisticated evaluation framework is needed, which could assess performance across multiple levels of biological and clinical abstraction. We propose a hierarchical evaluation strategy: 1) Low-level validation, focusing on molecular and cellular accuracy (e.g., protein folding, cellular localization); 2) Mid-level assessment, targeting tissue and organ-level simulations (e.g., structural deformation, fluid dynamics); 3) High-level clinical evaluation, encompassing systemic responses, disease progression, and treatment impact; 4) Longitudinal evaluation, monitoring how well the model adapts over time with new patient data.

The ability to perform both forward and inverse reasoning across these levels is essential. Moreover, as the model may generate novel hypotheses or out-of-distribution predictions (e.g., de novo structures or untested therapies), we recommend grounding these in biomedical priors or physical constraints to ensure plausibility[6]. The benchmarking framework itself should be adaptive and iterative, co-evolving with ongoing experimental findings and clinical feedback.

**5. Application and future directions**

As data volumes grow and model architectures evolve, the Virtual Eye has the potential to revolutionize many aspects of ophthalmology. **Figure 3** outlines several envisioned applications. While initial use cases may focus on improving medical education and streamlining clinical workflows, one of the most transformative applications lies in the realm of scientific research.

A mature Virtual Eye platform could serve as an in silico laboratory, enabling researchers to investigate complex biological mechanisms without immediate reliance on physical experiments[67]. By simulating ocular systems at multiple scales, the platform could help identify potential causal relationships underlying observed phenotypes with quantified uncertainty. This capability would not only allow for virtual validation of hypotheses but also foster hypothesis generation, guiding more targeted and efficient experimental designs. Through iterative interaction with the Virtual Eye, researchers could refine their understanding of disease mechanisms, optimize drug discovery pipelines, and even integrate with self-driving laboratories to automate and accelerate the scientific process[68].

In clinical practice, the Virtual Eye could fundamentally reshape how ophthalmic care is delivered. By comparing a patient's current status with their digital twin's predicted trajectory, clinicians could identify early deviations from healthy baselines, offering new opportunities for proactive screening and early intervention during routine eye exams. The system could also serve as a simulation tool, allowing clinicians to test different interventions before applying them in real life. In surgical contexts, such as cataract or refractive surgery, virtual rehearsals could help identify the optimal surgical strategy for a given patient. These capabilities are already beginning to emerge in clinical practice and are expected to expand rapidly. Beyond prediction and simulation, the Virtual Eye can act as a clinical decision-support system, analyzing thousands of similar cases to assist with risk stratification, diagnosis, and personalized treatment planning. By tailoring care to the individual rather than one-size-fits-all guidelines, it supports a shift toward precision ophthalmology.

Ultimately, the goal of the AI-powered Virtual Eye is not to replace clinicians but to augment their capabilities, enabling more proactive, precise, and personalized eye care. As the ecosystem surrounding the Virtual Eye matures, it is likely to become a cornerstone of both translational research and next-generation clinical practice.

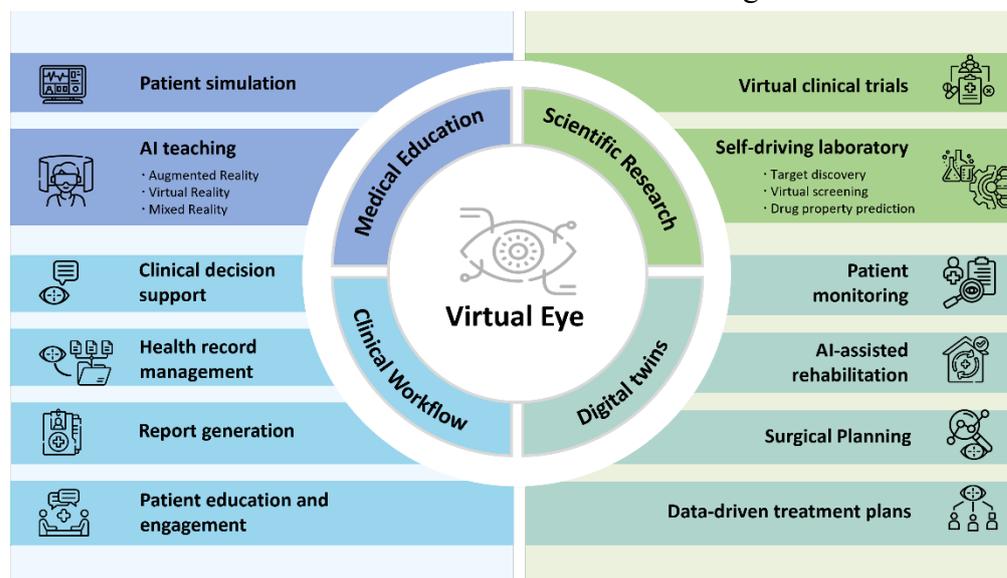

**Figure 3.** The application of virtual eye

## 6. Conclusion
The concept of an AI-powered virtual eye embodies a convergence of ophthalmology, computer science, mechanical engineering, and biology. In this perspective, we traced the evolution from early computational eye models to the current landscape shaped by AI, and outlined a forward-looking vision for a universal virtual eye. We presented a roadmap for realizing this vision, including data, modal architecture and interactive system. However, the need for virtual eye to process big amounts of data, achieve cross-context self-consistency, improve interpretability and reliability, and address ethical issues is critical for its broader application. Despite these challenges, the potential rewards are extraordinary. The virtual eye could usher in an era of precision ophthalmology and accelerate research as an in-silico laboratory. With the interdisciplinary collaboration across ophthalmologists, AI engineers, data scientists, ethicists, and policymakers, the AI-powered virtual eye can become a revolution and drive innovation in eye health management.

**Reference**

1   Burd, H. & Regueiro, R. Finite element implementation of a multiscale model of the human lens capsule. *Biomech Model Mechanobiol* **14**, 1363-1378 (2015).
2   Stewart, P. S., Jensen, O. E. & Foss, A. J. A theoretical model to allow prediction of the CSF pressure from observations of the retinal venous pulse. *Investigative Ophthalmology & Visual Science* **55**, 6319-6323 (2014).
3   Atchison, D. A. & Thibos, L. N. Optical models of the human eye. *Clin Exp Optom* **99**, 99-106 (2016). https://doi.org:10.1111/cxo.12352
4   Kuepfer, L., Fuellen, G. & Stahnke, T. Quantitative systems pharmacology of the eye: Tools and data for ocular QSP. *CPT: Pharmacometrics & Systems Pharmacology* **12**, 288-299 (2023).
5   Foster, W. J., Berg, B. W., Luminais, S. N., Hadayer, A. & Schaal, S. Computational Modeling of Ophthalmic Procedures: Computational Modeling of Ophthalmic Procedures. *American journal of ophthalmology* **241**, 87-107 (2022).
6   Bunne, C. *et al.* How to build the virtual cell with artificial intelligence: Priorities and opportunities. *Cell* **187**, 7045-7063 (2024). https://doi.org:10.1016/j.cell.2024.11.015
7   Thangaraj, P. M., Benson, S. H., Oikonomou, E. K., Asselbergs, F. W. & Khera, R. Cardiovascular care with digital twin technology in the era of generative artificial intelligence. *European Heart Journal* **45**, 4808-4821 (2024). https://doi.org:10.1093/eurheartj/ehae619
8   Johnson, G. T. *et al.* Building the next generation of virtual cells to understand cellular biology. *Biophysical Journal* **122**, 3560-3569 (2023). https://doi.org:10.1016/j.bpj.2023.04.006
9   Roberts, P. A., Gaffney, E. A., Luthert, P. J., Foss, A. J. E. & Byrne, H. M. Mathematical and computational models of the retina in health, development and disease. *Progress in Retinal and Eye Research* **53**, 48-69 (2016). https://doi.org:10.1016/j.preteyeres.2016.04.001
10  Gullstrand, A. in *Helmholtz's treatise on physiological optics, Vol. 1, Trans. from the 3rd German ed.*    301-358 (Optical Society of America, 1924).
11  Ding, X. *et al.* Time-Serial Evaluation of the Development and Treatment of Myopia in Mice Eyes Using OCT and ZEMAX. *Diagnostics (Basel)* **13** (2023). https://doi.org:10.3390/diagnostics13030379

12    Bakaraju, R. C., Ehrmann, K., Falk, D., Ho, A. & Papas, E. Optical performance of multifocal soft contact lenses via a single-pass method. *Optom Vis Sci* **89**, 1107-1118 (2012). https://doi.org:10.1097/OPX.0b013e318264f3e9

13    Inoue, M. *et al.* Quality of image of grating target placed in model of human eye with corneal aberrations as observed through multifocal intraocular lenses. *Am J Ophthalmol* **151**, 644-652 e641 (2011). https://doi.org:10.1016/j.ajo.2010.09.029

14    Li, Q. & Fang, F. Advances and challenges of soft contact lens design for myopia control. *Appl Opt* **58**, 1639-1656 (2019). https://doi.org:10.1364/AO.58.001639

15    Kawamorita, T. & Uozato, H. Modulation transfer function and pupil size in multifocal and monofocal intraocular lenses in vitro. *J Cataract Refract Surg* **31**, 2379-2385 (2005). https://doi.org:10.1016/j.jcrs.2005.10.024

16    Tognetto, D. *et al.* Analysis of the optical quality of intraocular lenses. *Invest Ophthalmol Vis Sci* **45**, 2682-2690 (2004). https://doi.org:10.1167/iovs.03-1024

17    Ackermann, R. *et al.* Optical side-effects of fs-laser treatment in refractive surgery investigated by means of a model eye. *Biomed Opt Express* **4**, 220-229 (2013). https://doi.org:10.1364/BOE.4.000220

18    Nguyen, B. A., Roberts, C. J. & Reilly, M. A. Biomechanical impact of the sclera on corneal deformation response to an air-puff: a finite-element study. *Front Bioeng Biotechnol* **6**, 210 (2019).

19    Braeu, F. A., Avril, S. & Girard, M. J. A. 3D growth and remodeling theory supports the hypothesis of staphyloma formation from local scleral weakening under normal intraocular pressure. *Biomech Model Mechanobiol* **23**, 2137-2154 (2024). https://doi.org:10.1007/s10237-024-01885-9

20    Zhao, Y. *et al.* Biomechanical analysis of ocular diseases and its in vitro study methods. *Biomed Eng Online* **21**, 49 (2022). https://doi.org:10.1186/s12938-022-01019-1

21    Boote, C. *et al.* Scleral structure and biomechanics. *Prog Retin Eye Res* **74**, 100773 (2020). https://doi.org:10.1016/j.preteyeres.2019.100773

22    Canning, C., Greaney, M., Dewynne, J. & Fitt, A. Fluid flow in the anterior chamber of a human eye. *Mathematical Medicine and Biology: a Journal of the IMA* **19**, 31-60 (2002).

23    Mauri, A. G. *et al.* Electro-fluid dynamics of aqueous humor production: simulations and new directions. *Modeling and Artificial Intelligence in Ophthalmology* **1**, 48-58 (2016).

24    Arciero, J. *et al.* Theoretical analysis of vascular regulatory mechanisms contributing to retinal blood flow autoregulation. *Investigative Ophthalmology & Visual Science* **54**, 5584-5593 (2013).

25    Jones, M., McElwain, D., Fulford, G., Collins, M. & Roberts, A. The effect of the lipid layer on tear film behaviour. *Bulletin of mathematical biology* **68**, 1355-1381 (2006).

26    Bhandari, A. Ocular Fluid Mechanics and Drug Delivery: A Review of Mathematical and Computational Models. *Pharm Res* **38**, 2003-2033 (2021). https://doi.org:10.1007/s11095-021-03141-6

27	Tripathy, K. C., Siddharth, A. & Bhandari, A. Image-based insilico investigation of hemodynamics and biomechanics in healthy and diabetic human retinas. *Microvasc Res* **150**, 104594 (2023). https://doi.org:10.1016/j.mvr.2023.104594

28	Zhou, Y. *et al.* A foundation model for generalizable disease detection from retinal images. *Nature* **622**, 156-163 (2023). https://doi.org:10.1038/s41586-023-06555-x

29	Qiu, J. *et al.* Development and validation of a multimodal multitask vision foundation model for generalist ophthalmic artificial intelligence. *NEJM AI* **1**, AIoa2300221 (2024).

30	Shi, D. *et al.* Eyefound: a multimodal generalist foundation model for ophthalmic imaging. *arXiv preprint arXiv:2405.11338* (2024).

31	Shi, D. *et al.* EyeCLIP: A visual-language foundation model for multi-modal ophthalmic image analysis. *arXiv preprint arXiv:2409.06644* (2024).

32	Koh, K., Tuuminen, R. & Jeon, S. Ultra-Widefield-Guided Swept-Source OCT Findings of Peripheral Vitreoretinal Abnormality in Young Myopes. *Ophthalmology* **131**, 434-444 (2024). https://doi.org:10.1016/j.ophtha.2023.11.009

33	Roorda, A. & Duncan, J. L. Adaptive optics ophthalmoscopy. *Annu Rev Vis Sci* **1**, 19-50 (2015). https://doi.org:10.1146/annurev-vision-082114-035357

34	Voigt, A. P. *et al.* Single-cell RNA sequencing in vision research: Insights into human retinal health and disease. *Prog Retin Eye Res* **83**, 100934 (2021). https://doi.org:10.1016/j.preteyeres.2020.100934

35	Choi, J. *et al.* Spatial organization of the mouse retina at single cell resolution by MERFISH. *Nat Commun* **14**, 4929 (2023). https://doi.org:10.1038/s41467-023-40674-3

36	Liu, W., Du, Z., Duan, Z., Li, L. & Shen, G. Neuroprosthetic contact lens enabled sensorimotor system for point-of-care monitoring and feedback of intraocular pressure. *Nat Commun* **15**, 5635 (2024). https://doi.org:10.1038/s41467-024-49907-5

37	Park, W. *et al.* In-depth correlation analysis between tear glucose and blood glucose using a wireless smart contact lens. *Nat Commun* **15**, 2828 (2024). https://doi.org:10.1038/s41467-024-47123-9

38	Chen, J. *et al.* Smartwatch Measures of Outdoor Exposure and Myopia in Children. *JAMA Netw Open* **7**, e2424595 (2024). https://doi.org:10.1001/jamanetworkopen.2024.24595

39	Mishra, T. *et al.* Pre-symptomatic detection of COVID-19 from smartwatch data. *Nat Biomed Eng* **4**, 1208-1220 (2020). https://doi.org:10.1038/s41551-020-00640-6

40	Shi, D. *et al.* Translation of color fundus photography into fluorescein angiography using deep learning for enhanced diabetic retinopathy screening. *Ophthalmology science* **3**, 100401 (2023).

41	Chen, R. *et al.* Translating color fundus photography to indocyanine green angiography using deep-learning for age-related macular degeneration

screening. *NPJ Digit Med* **7**, 34 (2024). https://doi.org:10.1038/s41746-024-01018-7

42  Wu, X. *et al.* FFA Sora, video generation as fundus fluorescein angiography simulator. *arXiv preprint arXiv:2412.17346* (2024).

43  Schuette, G., Lao, Z. & Zhang, B. ChromoGen: Diffusion model predicts single-cell chromatin conformations. *Sci Adv* **11**, eadr8265 (2025). https://doi.org:10.1126/sciadv.adr8265

44  Jumper, J. *et al.* Highly accurate protein structure prediction with AlphaFold. *Nature* **596**, 583-589 (2021). https://doi.org:10.1038/s41586-021-03819-2

45  Abramson, J. *et al.* Accurate structure prediction of biomolecular interactions with AlphaFold 3. *Nature* **630**, 493-500 (2024). https://doi.org:10.1038/s41586-024-07487-w

46  Lisanza, S. L. *et al.* Multistate and functional protein design using RoseTTAFold sequence space diffusion. *Nature biotechnology*, 1-11 (2024).

47  Shi, D. *et al.* Fundus2Globe: Generative AI-Driven 3D Digital Twins for Personalized Myopia Management. *arXiv preprint arXiv:2502.13182* (2025).

48  Eversberg, L. & Lambrecht, J. Combining synthetic images and deep active learning: Data-efficient training of an industrial object detection model. *Journal of Imaging* **10** (2024).

49  Taleb, A., Kirchler, M., Monti, R. & Lippert, C. in *Proceedings of the IEEE/CVF conference on computer vision and pattern recognition.*  20908-20921.

50  Chia, M. A. *et al.* Foundation models in ophthalmology. *The British Journal of Ophthalmology* **108**, 1341-1348 (2024). https://doi.org:10.1136/bjo-2024-325459

51  Nguyen, E. *et al.* Sequence modeling and design from molecular to genome scale with Evo. *Science* **386**, eado9336 (2024). https://doi.org:10.1126/science.ado9336

52  Hao, M. *et al.* Large-scale foundation model on single-cell transcriptomics. *Nat Methods* **21**, 1481-1491 (2024).

53  Lu, J. *et al.* Merge, ensemble, and cooperate! a survey on collaborative strategies in the era of large language models. *arXiv preprint arXiv:2407.06089* (2024).

54  Stahlschmidt, S. R., Ulfenborg, B. & Synnergren, J. Multimodal deep learning for biomedical data fusion: a review. *Briefings in bioinformatics* **23**, bbab569 (2022).

55  Ding, R., Lu, H. & Liu, M. DenseFormer-MoE: A Dense Transformer Foundation Model with Mixture of Experts for Multi-Task Brain Image Analysis. *IEEE Trans Med Imaging* **PP** (2025). https://doi.org:10.1109/TMI.2025.3551514

56  Gao, S. *et al.* Empowering biomedical discovery with AI agents. *Cell* **187**, 6125-6151 (2024). https://doi.org:10.1016/j.cell.2024.09.022

57  Li, B. *et al.* Mmedagent: Learning to use medical tools with multi-modal agent. *arXiv preprint arXiv:2407.02483* (2024).

58  Wang, M. *et al.* Common and rare fundus diseases identification using vision-language foundation model with knowledge of over 400 diseases. *arXiv preprint arXiv:2406.09317* (2024).

59  Navidi, Z. *et al.* MorphoDiff: Cellular Morphology Painting with Diffusion Models. *bioRxiv*, 2024.2012. 2019.629451 (2024).

60  Betzler, B. K. et al. Large language models and their impact in ophthalmology. *The Lancet Digital Health* **5**, e917-e924 (2023).

61  Yip, M. et al. Artificial intelligence meets medical robotics. *Science* **381**, 141-146 (2023). https://doi.org:10.1126/science.adj3312

62  Dwivedi, R. et al. Explainable AI (XAI): Core ideas, techniques, and solutions. *ACM Computing Surveys* **55**, 1-33 (2023).

63  Chen, R. J. et al. Algorithmic fairness in artificial intelligence for medicine and healthcare. *Nat Biomed Eng* **7**, 719-742 (2023). https://doi.org:10.1038/s41551-023-01056-8

64  Wang, Y., Liu, C., Zhou, K., Zhu, T. & Han, X. Towards regulatory generative AI in ophthalmology healthcare: a security and privacy perspective. *Br J Ophthalmol* **108**, 1349-1353 (2024). https://doi.org:10.1136/bjo-2024-325167

65  Yang, T., Ma, F., Qian, H. & Xu, B. AI-driven construction of digital cell model. *TIL* **2**, 100102 (2024). https://doi.org:10.59717/j.xinn-life.2024.100102

66  Chen, X. et al. Evaluating large language models and agents in healthcare: key challenges in clinical applications. *Intelligent Medicine* (2025). https://doi.org:https://doi.org/10.1016/j.imed.2025.03.002

67  Zhang, K. et al. Artificial intelligence in drug development. *Nat Med* **31**, 45-59 (2025). https://doi.org:10.1038/s41591-024-03434-4

68  Rapp, J. T., Bremer, B. J. & Romero, P. A. Self-driving laboratories to autonomously navigate the protein fitness landscape. *Nat Chem Eng* **1**, 97-107 (2024). https://doi.org:10.1038/s44286-023-00002-4

69  Thornburg, Z. R. et al. Fundamental behaviors emerge from simulations of a living minimal cell. *Cell* **185**, 345-360. e328 (2022).

70  Bhunia, S. S. & Saxena, A. K. Efficiency of Homology Modeling Assisted Molecular Docking in G-protein Coupled Receptors. *Curr Top Med Chem* **21**, 269-294 (2021). https://doi.org:10.2174/1568026620666200908165250

71  Jafarinia, H., Shi, L., Wolfenson, H. & Carlier, A. YAP phosphorylation within integrin adhesions: Insights from a computational model. *Biophys J* **123**, 3658-3668 (2024). https://doi.org:10.1016/j.bpj.2024.09.002

72  Keister, J., Cooper, E., Missel, P., Lang, J. & Hager, D. Limits on optimizing ocular drug delivery. *Journal of pharmaceutical sciences* **80**, 50-53 (1991).

73  Francoeur, M. L. KINETIC DISPOSITION AND DISTRIBUTION OF TIMOLOL IN THE RABBIT EYE.  (1984).

74  Miller, S. C., Himmelstein, K. J. & Patton, T. F. A physiologically based pharmacokinetic model for the intraocular distribution of pilocarpine in rabbits. *Journal of pharmacokinetics and biopharmaceutics* **9**, 653-677 (1981).

75  Luu, K. T. et al. Pharmacokinetic-pharmacodynamic and response sensitization modeling of the intraocular pressure-lowering effect of the EP4 Agonist 5-{3-[(2S)-2-{(3R)-3-hydroxy-4-[3-(trifluoromethyl) phenyl] butyl}-5-oxopyrrolidin-1-yl] propyl} thiophene-2-carboxylate (PF-04475270). *The Journal of pharmacology and experimental therapeutics* **331**, 627-635 (2009).

76  Karimi, A., Grytz, R., Rahmati, S. M., Girkin, C. A. & Downs, J. C. Analysis of the effects of finite element type within a 3D biomechanical model of a human optic nerve head and posterior pole. *Computer methods and programs in biomedicine* **198**, 105794 (2021).

77   Zhang, J., Qian, X., Zhang, H. & Liu, Z. Fluid-structure interaction simulation of aqueous outflow system in response to juxtacanalicular meshwork permeability changes with a two-way coupled method. *Computer Modeling in Engineering & Sciences* **116**, 301-314 (2018).

78   Zhang, H. *et al.* Stress–strain index map: A new way to represent corneal material stiffness. *Front Bioeng Biotechnol* **9**, 640434 (2021).

79   Knaus, K. R., Hipsley, A. & Blemker, S. S. The action of ciliary muscle contraction on accommodation of the lens explored with a 3D model. *Biomech Model Mechanobiol* **20**, 879-894 (2021).

80   Schwaner, S. A., Hannon, B. G., Feola, A. J. & Ethier, C. R. Biomechanical properties of the rat sclera obtained with inverse finite element modeling. *Biomech Model Mechanobiol* **19**, 2195-2212 (2020).

81   Listing, J. B. Dioptric des auges. *R. Wagners Handwörterb. D. Physiol* **4**, 451-504 (1851).

82   Emsley, H. H. *Visual Optics*.    (Hatton Press, 1946).

83   Rabbetts, R. B. *Bennett & rabbetts' clinical visual optics. 3. painos*.    (Oxford: Butterworth, 1998).

84   von Helmholtz, H. & Southall, J. P. C. *Helmholtz's Treatise on Physiological Optics*. (Optical Society of America, 1925).

85   LeGrand, Y. & ElHage, S. G. *Physiological optics*.    (Springer, 2013).

86   Drasdo, N. & Fowler, C. Non-linear projection of the retinal image in a wide-angle schematic eye. *The British journal of ophthalmology* **58**, 709 (1974).

87   Lotmar, W. Theoretical eye model with aspherics. *Journal of the Optical Society of America* **61**, 1522-1529 (1971).

88   Kooijman, A. C. Light distribution on the retina of a wide-angle theoretical eye. *Journal of the Optical Society of America* **73**, 1544-1550 (1983).

89   Navarro, R., Santamaria, J. & Bescós, J. Accommodation-dependent model of the human eye with aspherics. *J. Opt. Soc. Am. A* **2**, 1273-1280 (1985).

90   Thibos, L. N., Ye, M., Zhang, X. & Bradley, A. The chromatic eye: a new reduced-eye model of ocular chromatic aberration in humans. *Appl. Opt.* **31**, 3594-3600 (1992).

91   Liou, H.-L. & Brennan, N. A. Anatomically accurate, finite model eye for optical modeling. *J. Opt. Soc. Am. A* **14**, 1684-1695 (1997).

92   Guidoboni, G., Salerni, F., Repetto, R., Szopos, M. & Harris, A. Relationship between intraocular pressure, blood pressure and cerebrospinal fluid pressure: a theoretical approach. *Investigative Ophthalmology & Visual Science* **59**, 1665-1665 (2018).

93   Cassani, S., Arciero, J., Guidoboni, G., Siesky, B. & Harris, A. Theoretical predictions of metabolic flow regulation in the retina. *Modeling and Artificial Intelligence in Ophthalmology* **1**, 70-78 (2016).

94      Cassani, S., Harris, A., Siesky, B. & Arciero, J. Theoretical analysis of the relationship between changes in retinal blood flow and ocular perfusion pressure. *Journal of Coupled Systems and Multiscale Dynamics* **3**, 38-46 (2015).

95      Causin, P., Guidoboni, G., Malgaroli, F., Sacco, R. & Harris, A. Blood flow mechanics and oxygen transport and delivery in the retinal microcirculation: multiscale mathematical modeling and numerical simulation. *Biomech Model Mechanobiol* **15**, 525-542 (2016).

96      Liu, D. *et al.* Computational analysis of oxygen transport in the retinal arterial network. *Curr. Eye Res.* **34**, 945-956 (2009).

97      Aletti, M., Gerbeau, J.-F. & Lombardi, D. Modeling autoregulation in three-dimensional simulations of retinal hemodynamics. *Journal for Modeling in Ophthalmology* **1** (2015).

98      Kiel, J., Hollingsworth, M., Rao, R., Chen, M. & Reitsamer, H. Ciliary blood flow and aqueous humor production. *Progress in retinal and eye research* **30**, 1-17 (2011).

99      Diamond, J. M. & Bossert, W. H. Standing-gradient osmotic flow: a mechanism for coupling of water and solute transport in epithelia. *The Journal of general physiology* **50**, 2061-2083 (1967).

100     Repetto, R., Pralits, J. O., Siggers, J. H. & Soleri, P. Phakic iris-fixated intraocular lens placement in the anterior chamber: effects on aqueous flow. *Investigative Ophthalmology & Visual Science* **56**, 3061-3068 (2015).

101     Dvoriashyna, M., Repetto, R. & Tweedy, J. Oscillatory and steady streaming flow in the anterior chamber of the moving eye. *Journal of Fluid Mechanics* **863**, 904-926 (2019).

102     Abouali, O., Modareszadeh, A., Ghaffariyeh, A. & Tu, J. Numerical simulation of the fluid dynamics in vitreous cavity due to saccadic eye movement. *Medical Engineering & Physics* **34**, 681-692 (2012).

103     Tweedy, J. H., Pralits, J. O., Repetto, R. & Soleri, P. Flow in the anterior chamber of the eye with an implanted iris-fixated artificial lens. *Mathematical Medicine and Biology: A Journal of the IMA* **35**, 363-385 (2018).

104     Modarreszadeh, S., Abouali, O., Ghaffarieh, A. & Ahmadi, G. Physiology of aqueous humor dynamic in the anterior chamber due to rapid eye movement. *Physiology & behavior* **135**, 112-118 (2014).

105     Meskauskas, J., Repetto, R. & Siggers, J. H. Oscillatory motion of a viscoelastic fluid within a spherical cavity. *Journal of Fluid Mechanics* **685**, 1-22 (2011).

106     Repetto, R., Siggers, J. H. & Meskauskas, J. Steady streaming of a viscoelastic fluid within a periodically rotating sphere. *Journal of fluid mechanics* **761**, 329-347 (2014).

107     Stocchino, A., Repetto, R. & Cafferata, C. Eye rotation induced dynamics of a Newtonian fluid within the vitreous cavity: the effect of the chamber shape. *Physics in Medicine & Biology* **52**, 2021 (2007).

108     Repetto, R., Siggers, J. & Stocchino, A. Mathematical model of flow in the vitreous humor induced by saccadic eye rotations: effect of geometry. *Biomech Model Mechanobiol* **9**, 65-76 (2010).

109  Rossi, T. *et al.* Does the Bursa Pre-Macularis protect the fovea from shear stress? A possible mechanical role. *Experimental Eye Research* **175**, 159-165 (2018).

110  Bottega, W. J., Bishay, P. L., Prenner, J. L. & Fine, H. F. On the mechanics of a detaching retina. *Mathematical Medicine and Biology: A Journal of the IMA* **30**, 287-310 (2013).

111  Lakawicz, J. M., Bottega, W. J., Prenner, J. L. & Fine, H. F. An analysis of the mechanical behaviour of a detaching retina. *Mathematical Medicine and Biology: A Journal of the IMA* **32**, 137-161 (2015).

112  Guidoboni, G. *et al.* Intraocular pressure, blood pressure, and retinal blood flow autoregulation: a mathematical model to clarify their relationship and clinical relevance. *Investigative Ophthalmology & Visual Science* **55**, 4105-4118 (2014).

113  Winter, K. N., Anderson, D. M. & Braun, R. J. A model for wetting and evaporation of a post-blink precorneal tear film. *Mathematical Medicine and Biology: A Journal of the IMA* **27**, 211-225 (2010).

114  Braun, R. J., Gewecke, N. R., Begley, C. G., King-Smith, P. E. & Siddique, J. I. A model for tear film thinning with osmolarity and fluorescein. *Investigative Ophthalmology & Visual Science* **55**, 1133-1142 (2014).

115  Li, L., Braun, R., Henshaw, W. & King-Smith, P. Computed flow and fluorescence over the ocular surface. *Mathematical Medicine and Biology: A Journal of the IMA* **35**, i51-i85 (2018).

116  Peng, C.-C., Cerretani, C., Braun, R. J. & Radke, C. Evaporation-driven instability of the precorneal tear film. *Advances in colloid and interface science* **206**, 250-264 (2014).

117  Stapf, M., Braun, R. & King-Smith, P. Duplex tear film evaporation analysis. *Bulletin of Mathematical Biology* **79**, 2814-2846 (2017).

118  Zhong, L., Ketelaar, C., Braun, R., Begley, C. & King-Smith, P. Mathematical modelling of glob-driven tear film breakup. *Mathematical medicine and biology: a journal of the IMA* **36**, 55-91 (2019).

119  Heryudono, A. *et al.* Single-equation models for the tear film in a blink cycle: realistic lid motion. *Mathematical medicine and biology: a journal of the IMA* **24**, 347-377 (2007).

120  Aydemir, E., Breward, C. & Witelski, T. The effect of polar lipids on tear film dynamics. *Bulletin of mathematical biology* **73**, 1171-1201 (2011).

121  Sharma, A. & Ruckenstein, E. Mechanism of tear film rupture and formation of dry spots on cornea. *Journal of colloid and interface science* **106**, 12-27 (1985).

122  Nichols, J. J., Mitchell, G. L. & King-Smith, P. E. Thinning rate of the precorneal and prelens tear films. *Investigative Ophthalmology & Visual Science* **46**, 2353-2361 (2005).

123  Li, J. & Li, Y.-T. Extracting Inter-Protein Interactions Via Multitasking Graph Structure Learning. arXiv:2501.17589 (2025). <https://ui.adsabs.harvard.edu/abs/2025arXiv250117589L>.

124  Zhou, Z. *et al.* ProAffinity-GNN: A Novel Approach to Structure-Based Protein-Protein Binding Affinity Prediction via a Curated Data Set and Graph Neural

Networks. *J Chem Inf Model* **64**, 8796-8808 (2024). https://doi.org:10.1021/acs.jcim.4c01850

125   Réau, M., Renaud, N., Xue, L. C. & Bonvin, A. DeepRank-GNN: a graph neural network framework to learn patterns in protein-protein interfaces. *Bioinformatics* **39** (2023). https://doi.org:10.1093/bioinformatics/btac759

126   Truong-Quoc, C., Lee, J. Y., Kim, K. S. & Kim, D. N. Prediction of DNA origami shape using graph neural network. *Nat Mater* **23**, 984-992 (2024). https://doi.org:10.1038/s41563-024-01846-8

127   Rong, D., Zhao, Z., Wu, Y., Ke, B. & Ni, B. Prediction of Myopia Eye Axial Elongation With Orthokeratology Treatment via Dense I2I Based Corneal Topography Change Analysis. *IEEE Trans Med Imaging* **43**, 1149-1164 (2024). https://doi.org:10.1109/tmi.2023.3331488

128   Dai, Y., Gao, Y. & Liu, F. TransMed: Transformers Advance Multi-Modal Medical Image Classification. *Diagnostics (Basel)* **11** (2021). https://doi.org:10.3390/diagnostics11081384

129   Chen, X. *et al.* ChatFFA: An ophthalmic chat system for unified vision-language understanding and question answering for fundus fluorescein angiography. *iScience* **27**, 110021 (2024). https://doi.org:10.1016/j.isci.2024.110021

130   Chen, X. *et al.* EyeGPT for Patient Inquiries and Medical Education: Development and Validation of an Ophthalmology Large Language Model. *J Med Internet Res* **26**, e60063 (2024). https://doi.org:10.2196/60063

131   Chen, X. *et al.* FFA-GPT: an automated pipeline for fundus fluorescein angiography interpretation and question-answer. *npj Digital Medicine* **7**, 111 (2024). https://doi.org:10.1038/s41746-024-01101-z

132   Zhao, Z. *et al.* Slit Lamp Report Generation and Question Answering: Development and Validation of a Multimodal Transformer Model with Large Language Model Integration. *J Med Internet Res* **26**, e54047 (2024). https://doi.org:10.2196/54047

133   Chen, X. *et al.* ICGA-GPT: report generation and question answering for indocyanine green angiography images. *Br J Ophthalmol* **108**, 1450-1456 (2024). https://doi.org:10.1136/bjo-2023-324446

134   Chen, R. *et al.* EyeDiff: text-to-image diffusion model improves rare eye disease diagnosis. *arXiv preprint arXiv:2411.10004* (2024).

135   Mak, H. W. L., Han, R. & Yin, H. H. F. Application of Variational AutoEncoder (VAE) Model and Image Processing Approaches in Game Design. *Sensors* **23**, 3457 (2023).

Table 1. Representatives of the mechanistic and deep learning-based eye model

| Model type | Methodology | Domain target | Simulation use example |
|---|---|---|---|
| **Cellular & molecular** | Whole-cell kinetic model | Whole cell | Simulating cellular processes including metabolism, gene expression, and growth (e.g., JCVI-syn3A) [69] |
| | Molecular docking | Molecules (usually protein) | Predicting ligand-receptor interactions and binding affinity [70] |
| | Signaling network model | Intracellular signaling pathway | Dynamic simulation of signaling pathways (e.g. integrin-YAP) [71] |
| **Pharmacokinetic model** | Noncompartmental Model | Tear film and precorneal region | Predicting ocular bioavailability after topical eye drops [72] |
| | Classical Compartmental Model | Cornea, aqueous humor, vitreous | Drug distribution among ocular compartments [73] |
| | Physiologically Based Compartmental Model | Cornea, aqueous humor, ciliary body, iris | Detailed pilocarpine distribution simulation post-topical dosing [74] |
| | Population Model | Whole ocular system | Characterizing pharmacokinetic trends and variability across populations [75] |
| **Biomechanical model** | Finite Element Modeling | Optic nerve head; trabecular meshwork; cornea; lens and ciliary muscle | Simulations for glaucoma pathogenesis, refractive surgery outcome, keratoconus, and ocular accommodation [76-79] |
| | Inverse Finite Element Modeling | Sclera | Predicting scleral biomechanical properties related to glaucoma [80] |
| **Optical model** | Paraxial models/ finite models | Cornea, crystalline lens, retina | Clinical refraction analysis, visual assessment (scotomas/perimetry/aberration/MTF), surgical outcome prediction, IOL/contact lens design optimization [10,81-91] |

| Model type | Methodology | Domain target | Simulation use example |
|---|---|---|---|
| **Fluid-dynamical model** | Multi-dimensional (0D/1D/2D/3D) flow modeling | Retinal vasculature | Simulating blood flow dynamics, oxygen saturation in Glaucoma, AMD, DR, trabeculectomy; retinal oxygen saturation[24,92-97] |
| | Aqueous humor flow modeling | Anterior ocular segment | Simulating dynamics associated with glaucoma, refractive surgery, drug optimization[22,23,98-104] |
| | Vitreous humor modeling | Posterior ocular segment | Simulating dynamics associated with high myopia, retinal detachment, posterior vitreous detachment[105-111] |
| | Tear film dynamics | Ocular surface | Simulating dynamics associated with dry eye disease, meibomian gland dysfunction, post-LASIK tear instability, and Contact lens waering[25,112-122] |
| **Deep learning-based eye model** | GNN / CNN/ Transformer/ GAN LLM/ Diffusion Model/ VAE/ Foundation model | Text, image, video, 3D-shape | DNA/RNA/Protein structure/interaction prediction[123-126], disease classification/segmentation/prediction[127,128], question-answering[129-132], report generation[133], biology language processing, data synthesis and augmentation[42,47,134,135] |

**Table 2: Comparison of existing eye models and the concept of an AI-powered virtual eye**

| Characteristics | Stage 1: Mechanistic eye model | Stage 2: Deep-learning-based eye model | Stage 3: Universal virtual eye |
|---|---|---|---|
| **Underlying Principle** | rule-based, explicitly defined physical and biological equations | data-driven, learned statistical associations from large datasets | hybrid, integrates knowledge with data-driven learning and generative capabilities |
| **Model Flexibility** | low; highly specific to a particular problem | moderate; flexible for tasks within training distribution | high; general-purpose adaptability through interconnected foundation models |
| **Data Type** | usually single modality (e.g., molecules, optics, biomechanics, fluid dynamics) | primarily imaging-based; increasingly multimodal | fully multimodal integration: imaging, genomics, omics, clinical and environmental data |
| **Integration scale** | single-scale; (organ-level or molecular-level independently) | multi-scale, but limited in cross-scale molecular-to-organ integration | comprehensive multi-scale; molecules → pathways → cells → tissues → organs |
| **Predictive and generative capability** | limited beyond idealized assumptions | moderate within training distribution, reduced under data shifts | robust and adaptive across varying distributions and previously unseen scenarios |
| **Feedback** | minimal; static models updated manually | limited; periodic retraining with new data | continuous and dynamic adaptation through internal and external feedback loops |
| **Interpretability** | high | low to moderate | moderate |